%% file: med_arch2_wit.tex
\documentclass[11pt,english,australian]{revtex4-1}
\usepackage[T1]{fontenc}
\usepackage[utf8]{inputenc}
\usepackage{xcolor}
\usepackage{pdfcolmk}
\usepackage{amsmath}
\usepackage{graphicx}
\PassOptionsToPackage{normalem}{ulem}
\usepackage{ulem}

\makeatletter

\providecolor{lyxadded}{rgb}{0,0,1}
\providecolor{lyxdeleted}{rgb}{1,0,0}

\DeclareRobustCommand{\lyxsout}[1]{\ifx\\#1\else\sout{#1}\fi}

\usepackage{babel}
\usepackage{booktabs}
\usepackage{hyperref}
\usepackage{times}
\usepackage{adjustbox}

\makeatother

\usepackage{babel}
\begin{document}
\title{Perception of emergent epidemic of COVID-2019\foreignlanguage{english}{
/ SARS CoV-2} on the Polish Internet }
\author{Andrzej~Jarynowski}
\email{ajarynowski@interdisciplinary-research.eu}

\affiliation{\selectlanguage{english}%
Interdisciplinary Research Institute, Wroclaw, Poland}
\author{\selectlanguage{australian}%
Monika Wójta-Kempa}
\email{monika.wojta-kempa@umed.wroc.pl }

\affiliation{\selectlanguage{english}%
Department of Public Health, Faculty of Health Science, Wroclaw Medical
University, Poland}
\author{\selectlanguage{australian}%
Vitaly~Belik}
\email{vitaly.belik@fu-berlin.de}

\affiliation{\selectlanguage{english}%
System Modeling Group, Institute of Veterinary Epidemiology and Biostatistics,
Freie Universität Berlin}
\begin{abstract}
Problem: Due to the spread of SARS CoV-2 virus infection and COVID-2019
disease, there is an urgent need to analyze COVID-2019 epidemic perception
in Poland. This would enable authorities for preparation of specific
actions minimizing public health and economic risks. 

Methods: We study the perception of COVID-2019 epidemic in Polish
society using quantitative analysis of its digital footprints on the
Internet (on \emph{Twitter}, \emph{Google}, \emph{YouTube}, \emph{Wikipedia}
and electronic media represented by \emph{Event Registry}) from January
2020 to 12.03.2020 (before and after official introduction to Poland
on 04.03.2020). To this end we utilize data mining, social network
analysis, natural language processing techniques. Each examined internet
platform was analyzed for representativeness and composition of the
target group.

Results: We identified three temporal major cluster of the interest
before disease introduction on the topic COVID-2019: China- and Italy-related
peaks on all platforms, as well as a peak on social media related
to the recent special law on combating COVID-2019. Besides, there
was a peak in interest on the day of officially confirmed introduction
as well as an exponential increase of interest when the Polish government
``declared war against disease'' with a massive mitigation program.
From sociolingistic perspective, we found that concepts and issues
of threat, fear and prevention prevailed before introduction. After
introduction, practical concepts about disease and epidemic dominate.
We have found out that Twitter reflected the structural division of
the Polish political sphere. We were able to identify clear communities
of governing party, mainstream oppostition and protestant group and
potential sources of misinformation. We have also detected bluring
boundaries between comminities after disease introduction. 

Conclusions: Traditional and social media do not only reflect reality,
but also create it. Due to filter ``bubbles'' observed on Twitter,
public information campaigns might have less impact on society than
expected. For greater penetration, it might be necessary to diversify
information channels to reach as many people as possible which might
already be happening. Moreover, it might be necessary to prevent the
spread of disinformation, which is now possible due to the special
law on combating COVID-2019.
\end{abstract}
\maketitle
\selectlanguage{australian}%

\section*{Introduction}

\selectlanguage{english}%
Although a large part of the Polish population has heard about coronaviruses
for the first time a few weeks ago, in reality they face less dangerous
coronaviruses causing simple cold all the time. Only the emergence
of a novel strain from Wuhan gave the word ``Coronavirus'' a new
meaning. Before disease introduction, less than a half of surveyed
Poles believed that corona virus is the most important topic in the
second half of February 2020\foreignlanguage{australian}{ (\citet{ibris}).
}The disease was not detected in Poland until 03.03.2020, but this
topic was relatively important already before introduction and started
to drive media and social life after the disease introduction. During
opinion poll performed on 09.03.2020 and 10.03.2020, 63\% of respondents
reported, that this is a serious threat for Poland, 40\% that this
is a serious threat for their family and 73\% that this is a serious
threat for the Polish Economy (\citep{oko_2020}). We observe several
small peaks of interest on the Internet for Coronavirus in Poland\foreignlanguage{australian}{
during investigated period (since January 2020 till 11.03.2020),}
although the largest surge in interest occurs after the official confirmation
of the virus introduction to Poland on 04.03.2020 . The last day
of analysis 11.03.2020 was a day of pandemic declaration by WHO

Recently social media activities are being analyzed worldwide to better
understand perception and spread of diseases. This helps in some cases
to track the spread of diseases\foreignlanguage{australian}{~(\citet{ginsberg2009detecting,lu2018accurate,joshi2020harnessing})
with a higher precision than other methods. }The Internet is a good
ground for propagation of views often contradicting the current state
of medical knowledge\foreignlanguage{australian}{~}\citep{kata2012anti}.
Social media can serve as a valuable source of information as well
as disinformation\foreignlanguage{australian}{ }about the virus globally,
fueling panic and creating so-called infodemic (\citep{natasza}),
at unprecedented speed and massively desrupting entire countries such
as Italy\foreignlanguage{australian}{~(\citet{guardian}).} Propaganda
and persuasion techniques are widely used on the Internet easily reaching
certain target groups susceptible to conspiracy theories and effectively
polarizing societies, possibly due to the interference by foreign
intelligence\foreignlanguage{australian}{~}\citep{lightfoot2017political}.
Panic-related behaviors accompanying a virus outbreak\foreignlanguage{australian}{~}\citep{gonsalves2014panic}
are influencing the epidemic spread~(\citet{perra2011towards,fenichel2011adaptive,wang2015coupled}),
with the Internet being the main mediating mechanism. This could lead
to destructive collective behavior, such as e.g.\foreignlanguage{australian}{~}xenophobia
against people from affected countries. This is reflected on the Internet
as symbolic violence and hate – quote from Twitter: \textquotedbl Chinese
should be banned from entering our country\textquotedbl .
\selectlanguage{australian}%

\subsection*{Methodology}

\selectlanguage{english}%
In this study, by quantitative analysis of digital traces on the Internet
(e.g. social media), we concetrate on following key dimensions as
the number and nature of social media events such as information queries
and deploy of social network analysis.

\selectlanguage{australian}%
Up to our knowledge there were no previous studies quantitatively
linking the Internet activities and risk perception of infectious
diseases in Poland (\citep{samaras2020syndromic,nuti2014use}). Thus
the present study is a first exploratory attempt filling this gap
after prelimiary research om Coronovirus perception in Poland before
disease introduction (\citep{jarynowski2020percepcja}).

\selectlanguage{english}%
We primarily analyze quantitative digital footprint data on the Internet
from January 2020 to 11.03.2020, including their representativeness.
The number of internet users in Poland in January 2020 was 28.1 million\foreignlanguage{australian}{
(\citet{pbi_20}) and 28.6 milion in 2019 (\citet{pbi_19})}. Internet
covers 85\% of the total literate population in Poland. Thus, the
passive representativeness of the Internet is relatively high, but
active (own content creation) is biased towards younger age groups
and women. The former group has a high activity on the Internet, e.g.
an average Polish teenager spends about 5 hours a day on the Internet\foreignlanguage{australian}{
(\citet{tanas2017raport}))} and the later group is responsible for
generation of up to 85\% of health-related content in social media(\foreignlanguage{australian}{(\citet{pbi_20}}).
Over 99\% of young Polish women use the Internet\foreignlanguage{australian}{
(\citet{jarynowski2018choroby})). }

\selectlanguage{australian}%
As a keyword in queries of web services (\citep{nuti2014use}) we
chose a colloqial term ``Koronawirus''/''Coronavius'' due to its
penetration in the society. Other related keywords in use are much
less popular, except for Wikipedia, where medical term SARS-CoV-2
was choosen. . 

\selectlanguage{english}%
In our study each considered internet platform is described separately
and has its own specific bias. Data analysis can be biased e.g. due
to content presenting algorithms on the media platforms. For instance
technological giants Google, Twitter and Facebook are supposed to
implement fact verification algorithms to filter out false informations.
Being aware of this, computational techniques of social sciences\foreignlanguage{australian}{
(\citet{Obliczeniowe}),} despite some disadvantages and their exploratory
nature, provide us with an opportunity to analyze a huge amount of
digital footprint data at low cost and in short time.
\selectlanguage{australian}%

\section*{Google trends}

\begin{figure}
\includegraphics[width=0.95\textwidth]{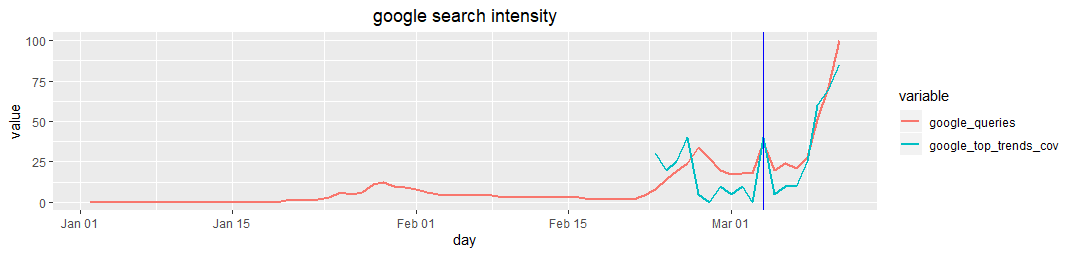}

\caption{\label{goog_trends} \foreignlanguage{english}{The intensity of searched
queries with the word \textquotedbl corona virus\textquotedbl /``Koronawirus''
in Polish Google (01.01-11.03.20) and percentage of ``coronavius''
related queries in top trend topics (23.01-11.03.20) both generated
using Google Trend tool. Disease introduction marked with the blue
vertical line.}}
\end{figure}

\selectlanguage{english}%
Google has 95\% share among Polish Internet users with over 8 billion
entries per month and is the undisputed leader on the Polish Internet
market\foreignlanguage{australian}{ (\citet{pbi_20}).} Interest in
novel Coronavirus on Google can be measured by the number of queries\foreignlanguage{australian}{~{[}Fig.
\ref{goog_trends}{]}}. According to Google there were $2\cdot10^{5}$
searches of term ``Koronawirus'' monthly. However, it is calculated
based on historical data. Thus there were dozens of thousand daily
searches in the late February 2020. 

\selectlanguage{australian}%

\selectlanguage{english}%
The subject of the corona virus occurred recently after the outbreak
in Wuhan. It is important to note, that until disease introduction
to Poland there were no Coronavirus related searches in top 25 Google
queries at all. Although ``coronavirus'' related queries were observed
in top trends before the introduction of the disease to Poland, they
start to dominate top queries only after the massive mitigation measures
were taken such as school/university and boarder closures around 09-12.03.20\foreignlanguage{australian}{
{[}Fig. \ref{goog_trends}{]}}. Prior to the disease introduction
two phases of interest can be distinguished\foreignlanguage{australian}{~{[}Fig.
\ref{goog_trends}{]}:}
\begin{enumerate}
\item from the end of January and beginning of February when the epidemic
was announced and confirmed in China. We see a small peak around 25.01.20
(e.g. death of Liang Wudong) and around 29.01.20 (e.g. first case
in Germany);
\item from the end of February till beginning of March (when the number
of infections rapidly increased in Italy). We see a clear peak around
27.02.20 (e.g. fake news about possible introduction of the disease
to Poland \citep{wyborcza}). 
\end{enumerate}
After the first confirmed case in the country we can see a peak on
that day (04.03.20) and substantial growth after important measures
were implemented by Polish authorities (09-11.03).

\begin{figure}
\includegraphics[width=0.95\textwidth]{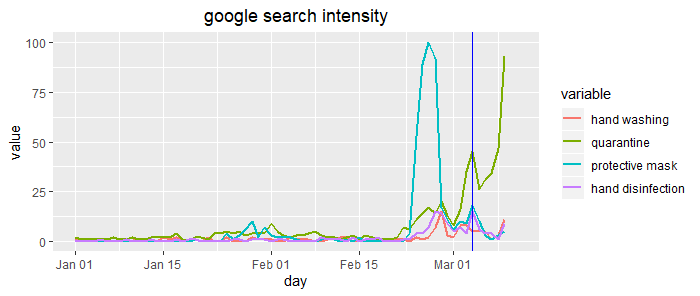}

\caption{\label{goog_maski} \foreignlanguage{english}{The intensity of queries
with the phrases ``quarantine'', \textquotedbl protective mask\textquotedbl ,
\textquotedbl hand washing\textquotedbl , \textquotedbl hand disinfection\textquotedbl{}
(} ``kwarantanna'' ``maseczka ochronna'', ``mycie rąk'', ``dezynfekcja
rąk'')\foreignlanguage{english}{ in Polish Google (01.01-10.03/2020)
generated using the Google Trend tool. Disease introduction marked
with the blue vertical line.}}
\end{figure}

People are looking for information on further epidemiological topics
related to infection and epidemics\foreignlanguage{australian}{ as
well {[}Fig. \ref{goog_maski}{]}. }It should be noted that professional
vocabulary such as \textquotedbl hand hygiene\textquotedbl{} practically
does not appear in queries (below the ``noise threshold'' compared
to other epidemiological terms\foreignlanguage{australian}{ {[}Fig.
\ref{goog_maski}}{]}). Issues related to hand hygiene have a peak
in late February and beginning of March. Masks search had their peak
of popularity at the end of February and the lack of increase in popularity
in March (compared to other epidemiological queries \foreignlanguage{australian}{{[}Fig.
\ref{goog_maski}}{]}) may be due to the success of information campaigns
on their alleged low effectiveness or simply due to the lack of the
masks on the market.

Search activity for information on infection in Poland is still much
lower than in countries with high global mobility\foreignlanguage{australian}{
(\citet{lai2020assessing}) }and already confirmed cases of infection\foreignlanguage{australian}{
.} Such a peripheral location of Poland (among others) lead to the
first confirmed case introduced by land rather than by air\foreignlanguage{australian}{
(\citet{inter}).}

\selectlanguage{australian}%

\begin{figure}
\includegraphics[width=0.45\textwidth]{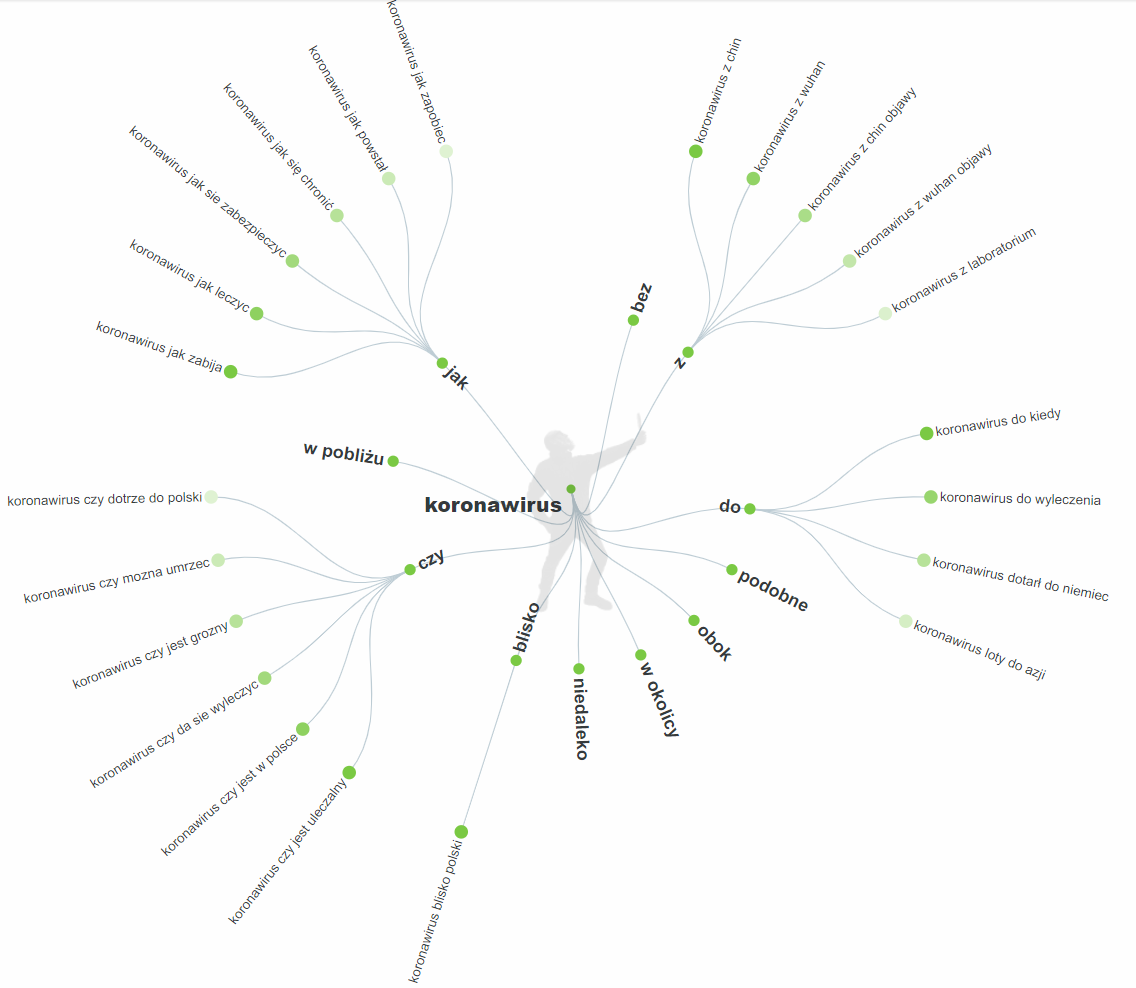}\includegraphics[width=0.45\textwidth]{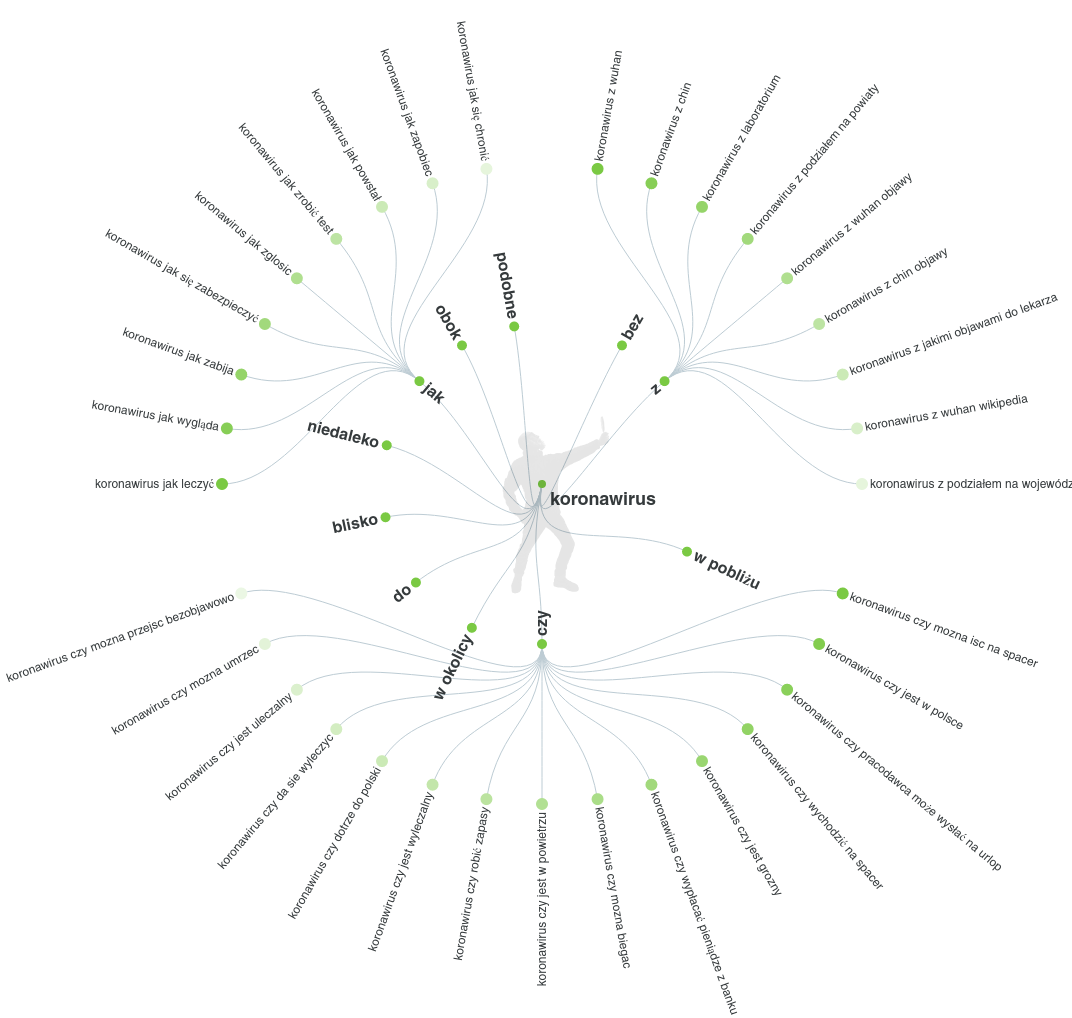}

\caption{\label{ask_public} \foreignlanguage{english}{Semantic web (prepositions)
of the noun \textquotedbl corona virus\textquotedbl{} in Google search
engine before introduction (snapshot on 29.02.20) {[}left{]} and after
introduction (snapshot on 12.03.20) {[}right{]} generated using the
Answer the Public tool for Poland and Polish} \foreignlanguage{english}{language}
(\citet{ask_public}).}
\end{figure}

\selectlanguage{english}%
In the corresponding semantic networks we recognize the most common
co-occurring phrases in the search engine together with the word \textquotedbl Coronavirus\textquotedbl\foreignlanguage{australian}{
{[}Fig. \ref{ask_public}{]}.} Such a network contains information
on how the predictor - the corona virus noun - relates to its arguments
in a query phrase\foreignlanguage{australian}{ (\citep{maziarz2016plwordnet}).}

\selectlanguage{australian}%

Before virus introduction, we observed {[}Fig. \ref{ask_public}{]},
\foreignlanguage{english}{that most often the questions are associated
with a threat, e.g. if \emph{it is / will reach / to Poland / close
to Poland; can one die / how it kills}} (czy jest / dotrze w / do
Polski(ce) / blisko Polski; czy można umrzeć / jak zabija).\foreignlanguage{english}{
Second level searches concerns prevention, e.g. \emph{how to prevent
/ guard / protect yourself}} (jak zapobiec / chronić się / zabezpieczyć
się).\foreignlanguage{english}{ In addition, there are third level
threads such as symptoms, history, or restrictions. The aspect of
geographical proximity is also very important, thus the terms \emph{nearby},
\emph{near} next to dominant semantic field around the word \textquotedbl Coronavirus\textquotedbl . }

\selectlanguage{english}%
After virus introduction, there is domination of practical questions
e.g. \emph{how to do the test / when to call a doctor / how to treat
/ asymptomatic course / can you go for a walk / is it in air/ do you
do shopping / leave from the employer / map} (jak zrobić test/ kiedy
dzwonić po lekarza/ jak się\ leczyć/ przebieg bezobjawowy/ czy można
iść na spacer/ czy jest w powietrzu/ czy robić zasapy/ urlop od pracodawcy/
mapy).
\selectlanguage{australian}%

\section*{Wikipedia}

\begin{figure}
\includegraphics[width=0.95\textwidth]{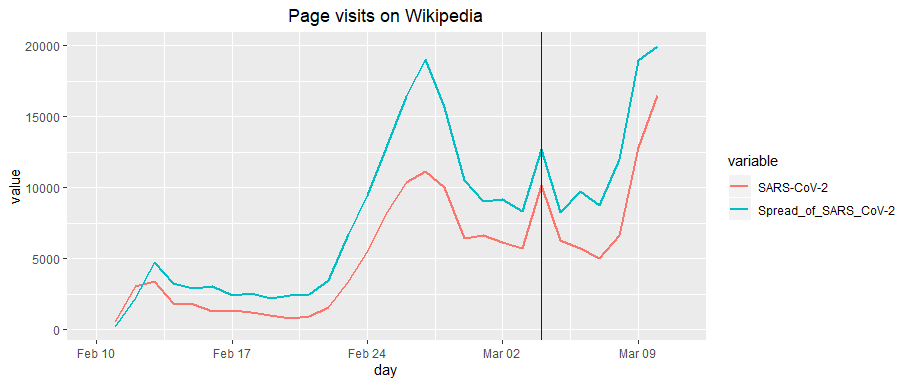}

\caption{\foreignlanguage{english}{The number of views on the article \textquotedbl Spread of SARS-CoV-2
virus infection\textquotedbl /\foreignlanguage{australian}{ ``Szerzenie
się zakażeń wirusem SARS-CoV-2''} and\foreignlanguage{australian}{
``SARS-Cov-2'' (10.02-10.03.2020) on Wikipedia. }Disease introduction
marked with the blue vertical line.\foreignlanguage{australian}{\label{wiki}}}}
\end{figure}

\selectlanguage{english}%
Wikipedia traffic is another indicator of the social activity. Wikipedia
has an Internet coverage of 57\% with over 350 million entries per
month among Polish Internet users\foreignlanguage{australian}{ (\citet{pbi_20}).}
There is a significant overepresentation of users with tertiary education
inhabiting big cities (affinity index>120 (\citep{pbi_wiki})) We
looked at the history of page views on discussions around the articles\foreignlanguage{australian}{
``SARS-CoV-2'' (\citep{wiki_SARS-CoV-2}) }and \textquotedbl Spread
of virus infection SARS-CoV-2\textquotedbl\foreignlanguage{australian}{/``Szerzenie
się zakażeń wirusem SARS-CoV-2'' (on 11.03.2020 this page was taken
down and new page abou pandemic appeared in this place (\citep{wiki_szerzenie}))
in Polish Wikipedia {[}Fig. \ref{wiki}{]}. }

\selectlanguage{australian}%
\foreignlanguage{english}{Before introduction, we see a growing trend
in the number of queries with a small peak around 13.02.20 (which
does not appear in other media and is related probably to giving a
new name to the virus and the disease) and a clear peak around 27.02.20
(e.g. a fake news about disease in Poland (\citep{wyborcza})). The
first days of March are characterized by a slight decline in interest,
perhaps due to the saturation of knowledge of basic definitions about
the disease in the society. After the first confirmed case, there
is a small peak around this day and a slight growth during actual
epidemic and massive mitigation strategies. On 04.03.20, a new page
for epidemic spread in Poland was launched. }

\selectlanguage{english}%
The intensive discussions between editors in Wikipedia concern, among
other, the effectiveness of protective masks or the reliability of
data from the PRC (People's Republic of China). No data is available
before 10.02.2020, due to changes in the title of the articles together
with the name of the virus and disease by WHO.
\selectlanguage{australian}%

\section*{Event Registry}

We choose EventRegistry (\citet{Event}) \foreignlanguage{english}{as
a traditional media search engine because it has a large range of
online magazines representing various political sites. In addition,
it gives priority to the digital versions of other broadcasting channels,
including television, radio or newspapers. Between January 31 and
March 11, 10755 (before 4603 and 6152 after introduction) representative
articles were selected (the non-systematic sampling method was applied).}

\begin{figure}
\hfill{}\includegraphics[width=0.8\textwidth]{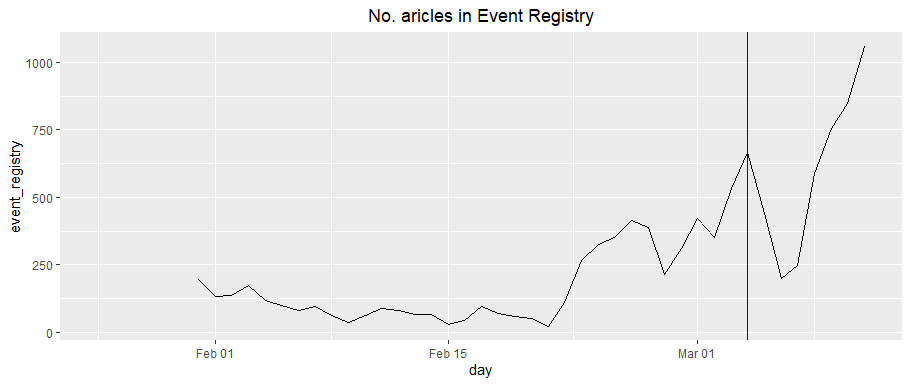}\hfill{}

\caption{\label{event_dynamika}\foreignlanguage{english}{Number of articles
in time 31.01-11.03 (generated using the Event Registry tool). Disease
introduction marked with the blue vertical line.}}
\end{figure}

\selectlanguage{english}%
In the traditional media the weekly seasonality of the articles and
3 peaks of interest: at the end of January, the second half of February
and the beginning of March\foreignlanguage{australian}{ can be cleary
seen~{[}Fig. \ref{event_dynamika}{]}. Counts of news article seems
to coindence with all peaks observed at other platforms and in some
cases as peak on 26.02, it aheads all of other platforms.}

\selectlanguage{australian}%

\section*{Twitter}

\selectlanguage{english}%
Twitter in Poland has relatively low popularity (\textasciitilde 3
million registered users or less than 8\% of the population) and is
mainly used by expats, journalists and politicians\foreignlanguage{australian}{
(\citet{sot_T}). }However, Twitter provides an API for data acquisition
available to general public almost for free. This allowed us to analyze
not only content of tweets, but also their context (location, retweeting,
commenting, etc.). A lot of interest in infection can be seen on Twitter
in Polish (210182 tweets with \#Koronawirus in about 50 days).

\selectlanguage{australian}%
\begin{figure}
\hfill{}\includegraphics[width=0.8\textwidth]{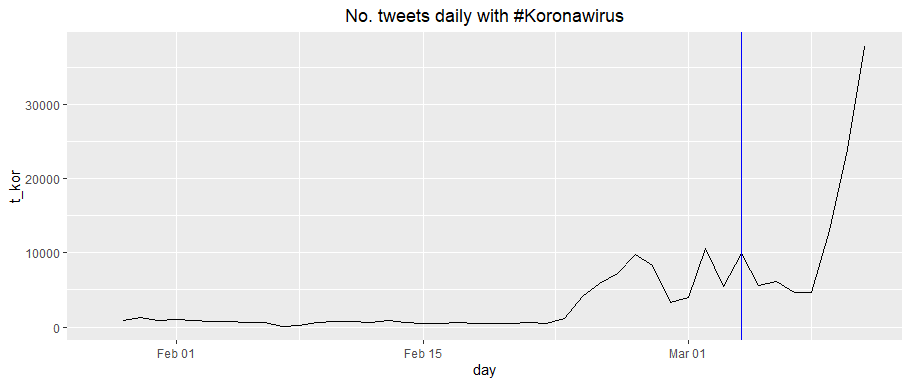}\hfill{}

\caption{\label{zli_twi}\foreignlanguage{english}{ Number of tweets per day
with the Koronawirus hashtag in Polish language} (28.01-11.03.2020).\foreignlanguage{english}{
Disease introduction marked with the blue vertical line}}
\end{figure}

There is not so much attention on Twitter until late February 2020.
There are peaks on 27.02.20 (fake news about possible case in Poland),
02.03.20 (discussion about special anti-COVID-2010 act (\citep{sejm})),
04.03.20 (disease introduction) and huge increase in interest on 9-11.03.20
(massive mitigation strategies implemented).

\begin{figure}
\includegraphics[width=0.8\textwidth]{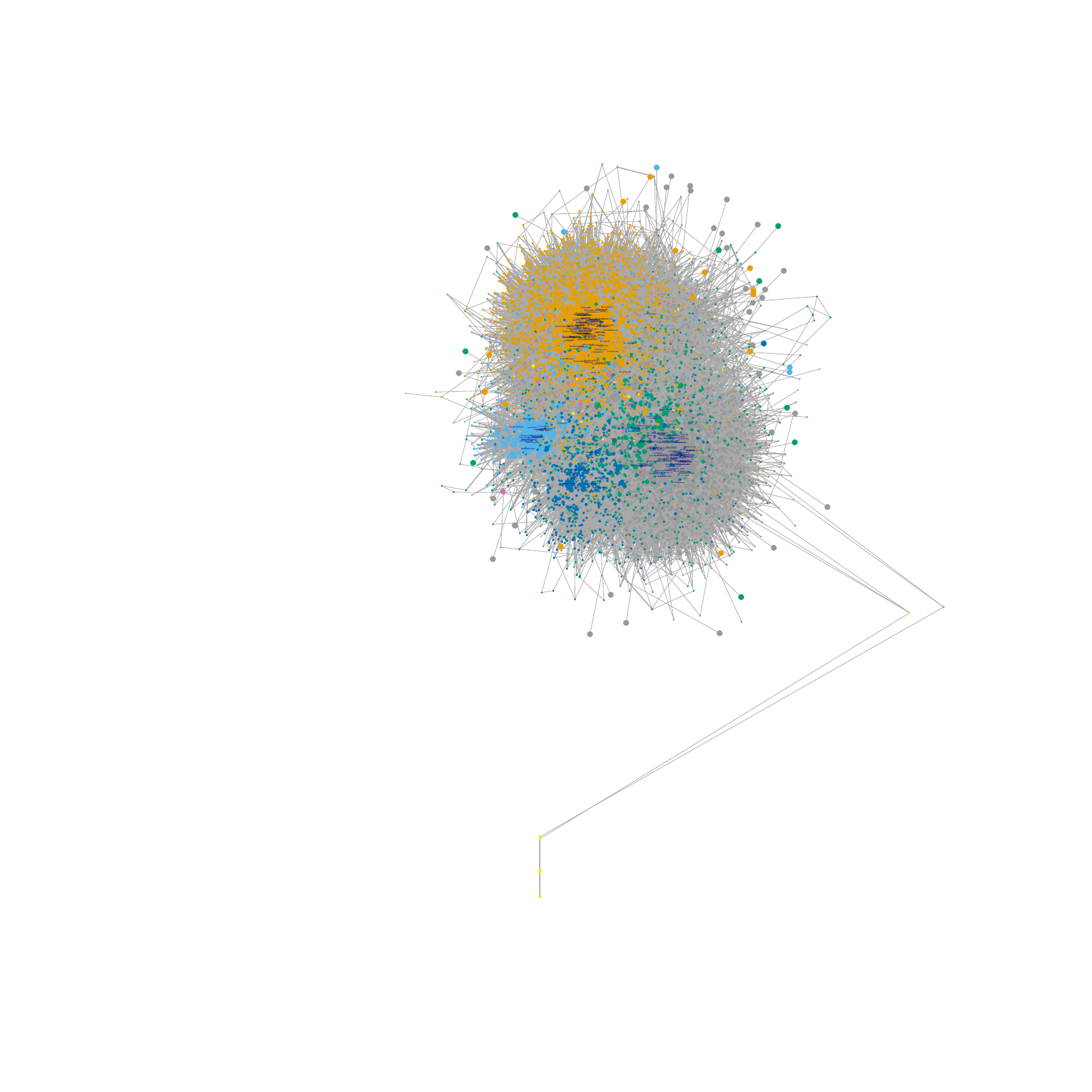}

\caption{\label{net_twi}\foreignlanguage{english}{A network of Twitter accounts
(vertices) connected by retweets (edges) with the Koronawirus hashtag.
Colorcode: gray - the ruling camp, orange - the opposition, light
blue - the Protestant group, dark blue - far right }(28.01-11.03.2020).
\foreignlanguage{english}{This network only shows accounts that have
generated at least 3 tweets and connections that represent at least
2 retweets, labels are provided fo 100 most central (weighted degree
centrality) .}}
\end{figure}

To apply Social Network Analysis\foreignlanguage{english}{~(\citet{wasserman1994social})}
methods to the Twitter data, we build a network with vertices representing
Twitter accounts and edges representing retweets\foreignlanguage{english}{~}(\citet{modelowanie}).
The network\foreignlanguage{english}{ revealed various connections
(social impact, trust, friendship, etc.) between accounts being social
actors and the characteristics of the actors (political affiliation,
views, etc.). An unsupervised weighted Louvain algorithm~(\citet{blondel2008fast})
for community analysis was used and the vertex color denotes the community
it belongs to. Retweet network} {[}Fig.\foreignlanguage{english}{~}\ref{net_twi}{]}
\foreignlanguage{english}{shows how the discourse is divided into
the ruling party (gray), opposition (orange) and the protestant religious
and political group / ,,Idz pod prad'' (light blue).}

\selectlanguage{english}%
In addition, further discourse participants were identified\foreignlanguage{australian}{
{[}Fig.~\ref{net_twi}{]}.} The subject of \textquotedbl Coronavirus\textquotedbl{}
in Poland has a conflict-generating potential. As a consequence a
dispute has emerged between the ruling party promoting information
content and confirming the belief that the Polish state is well prepared
to fight the virus (gray cluster), and the opposition negating its
ability to fight the virus\foreignlanguage{australian}{ {[}Fig.~\ref{net_twi}{]}.
}However, the most distinglished discourse is repesented in the cores
of the net \foreignlanguage{australian}{{[}Fig.~\ref{net_twi}{]}
and boundaries between communities are more blured than it was before
confirmation of the first case in Poland (\citep{jarynowski2020percepcja}).
It could mean, that (at least in this weekly resolution) for an average
Twitter user, mechanisms of community building (for example to help
neigbors and support healthworkers) are less politically driven as
it was before.}

For example, Twitter accounts classified already as potentially belonging
to the so-called trolls (which in other studies were classified to
the extreme right in the context of elections to the European Parliament\foreignlanguage{australian}{
(\citet{oko}), or to the far-left side in the context of the African
Swine Fever epidemic (\citet{jarynowski2019african}), p}romoted content
in the buffer area (attacking both the ruling party and the mainstream
opposition). \foreignlanguage{australian}{}
\begin{table}
\selectlanguage{australian}%
\input{img_dir/table211.tex}

\caption{Counts of most frequent words without Stop words during 28.01-11.03.2020.\label{slowa_twi}}
\selectlanguage{english}%
\end{table}

We also looked on the most frequent words after steming and excluding
Stop words\foreignlanguage{australian}{~}{[}Tab.\foreignlanguage{australian}{~}\ref{slowa_twi}{]}.
There are mainly words related to topics around health, but with politicical
context as well, due to the political and journalistic bias of Twitter.
In comparition to situation before the first confrimed case (\citep{jarynowski2020percepcja}),
there is less politics and there less fear related concepts.
\selectlanguage{australian}%

\section*{Youtube}

\begin{figure}
\includegraphics[width=0.9\textwidth]{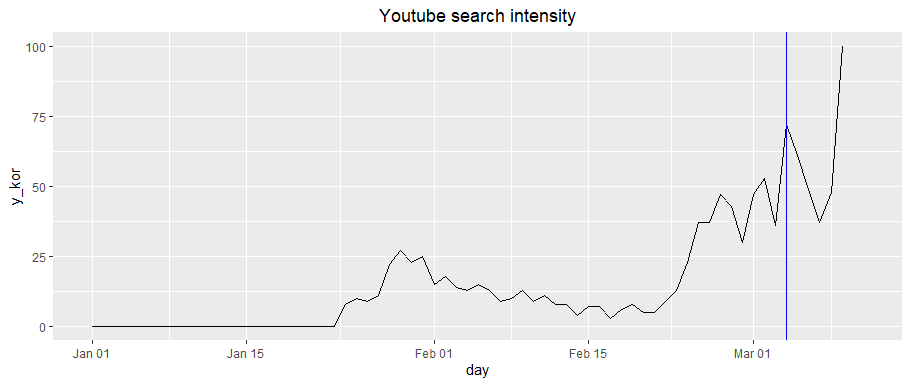}

\caption{\label{youtube_trends} \foreignlanguage{english}{The intensity of
queries for the word \textquotedbl Koronawirus\textquotedbl{} on
Youtube (01.01-11.03.20) generated using the Google Trend \uline{}
too}l. \foreignlanguage{english}{Confirmed disease introduction is
marked by the blue vertical line.}}
\end{figure}

\selectlanguage{english}%
Youtube has a share among Polish Internet users at the level of 68\%
with about 700 million entries per month\foreignlanguage{australian}{
(\citet{pbi_20}). }In addition, streams from the mobile app should
be taken into account, as it is most popular app on Poles' smartphones\foreignlanguage{australian}{
(\citet{pbi_20}). }For our analysis, we selected videos on the main
subject of \textquotedbl corona virus\textquotedbl\foreignlanguage{australian}{
(\citep{Youtube}) using keyword search.} One observes multiple peaks
(25.01, 31.01, 27.02, 2.03, 04.03, and 9-11.03.20) on Youtube similar
to all other social media\foreignlanguage{australian}{ {[}Fig. \ref{youtube_trends}{]}.}

\selectlanguage{australian}%

\section*{Other Media}

\selectlanguage{english}%
The fastest growing profile in January and February 2020 in the entire
Polish Facebook was \textquotedbl Conflicts and global disasters\textquotedbl{}
(,,Konflikty i katastrofy światowe'') which gained over 120 thousand
followers in one month correlates with recent increased activity related
to corona virus information. The most popular post in the Vlog category
in January 2020 was a video material titled \textquotedbl Wuhan market\textquotedbl{}
on SA Wardęga's profile (@sawardega), which was eventually marked
as containing false information\foreignlanguage{australian}{~(\citet{sot_F}). }

There are other social platforms not covered by this study such as
Instagram (a popular platform among teenagers with affinity index>120
in this age category\foreignlanguage{australian}{~}(\citep{pbi_instagram})).
The topic of Coronavius was not so popular at Polish Instagram until
the disease introduction. The most popular hashtag after introduction
of the virus ``odwolajcieszkolyxd'' and is related to school closures.

For a better topic coverage blogs (e.g. blog Media@jesion gathers
hundreds thousands entries by day\foreignlanguage{australian}{~}(\citep{blog})
and user comments in Internet media presents addition reach information
source. E.g. articles from media such as wp.pl, onet.pl, interia.pl
have billions entries monthly and their articles on Coronavirus reaches
on average dozens of thousands entries with hundreds of comments (\citep{pbi_20}).

\selectlanguage{australian}%

\section*{Comparison of different platforms}

\begin{figure}
\hfill{}\includegraphics[width=0.7\textwidth]{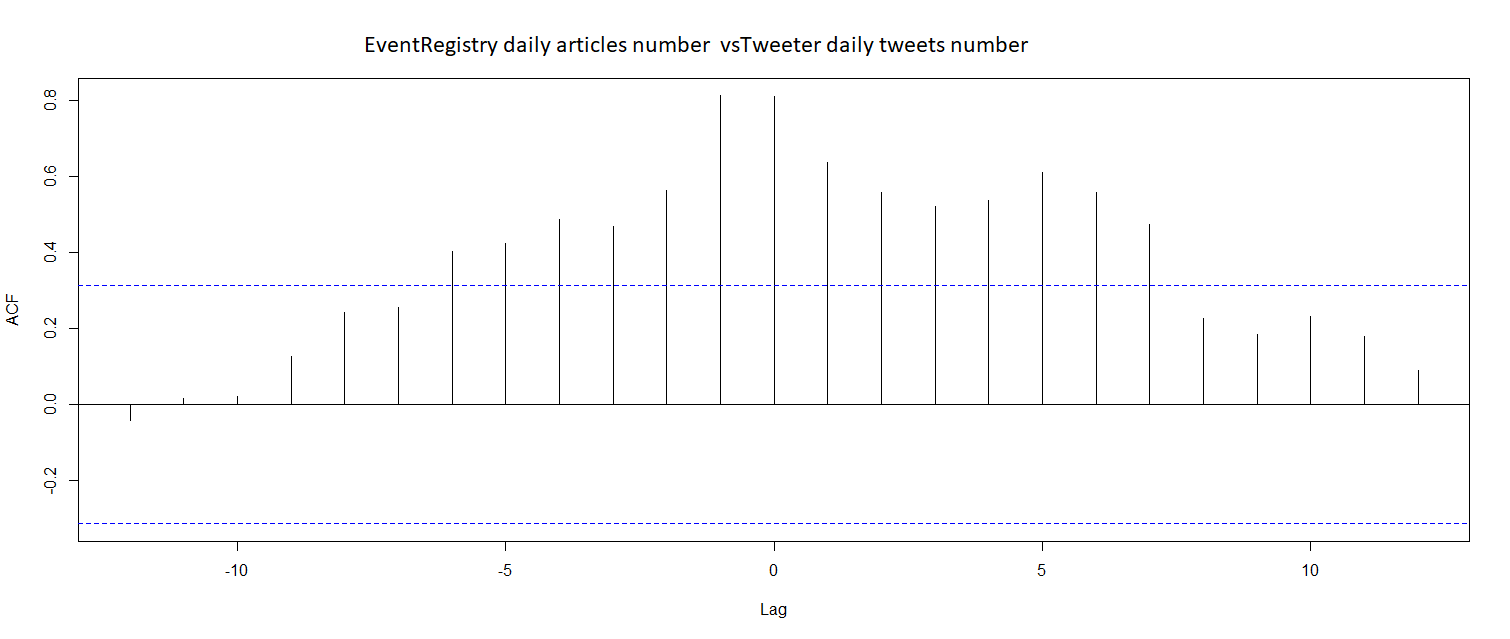}\hfill{}

\caption{\label{lagged} Lagged (in days) correlation between daily series
of article counts from Event Registry and Tweets numbers.}
\end{figure}

\begin{figure}
\includegraphics[width=0.99\textwidth]{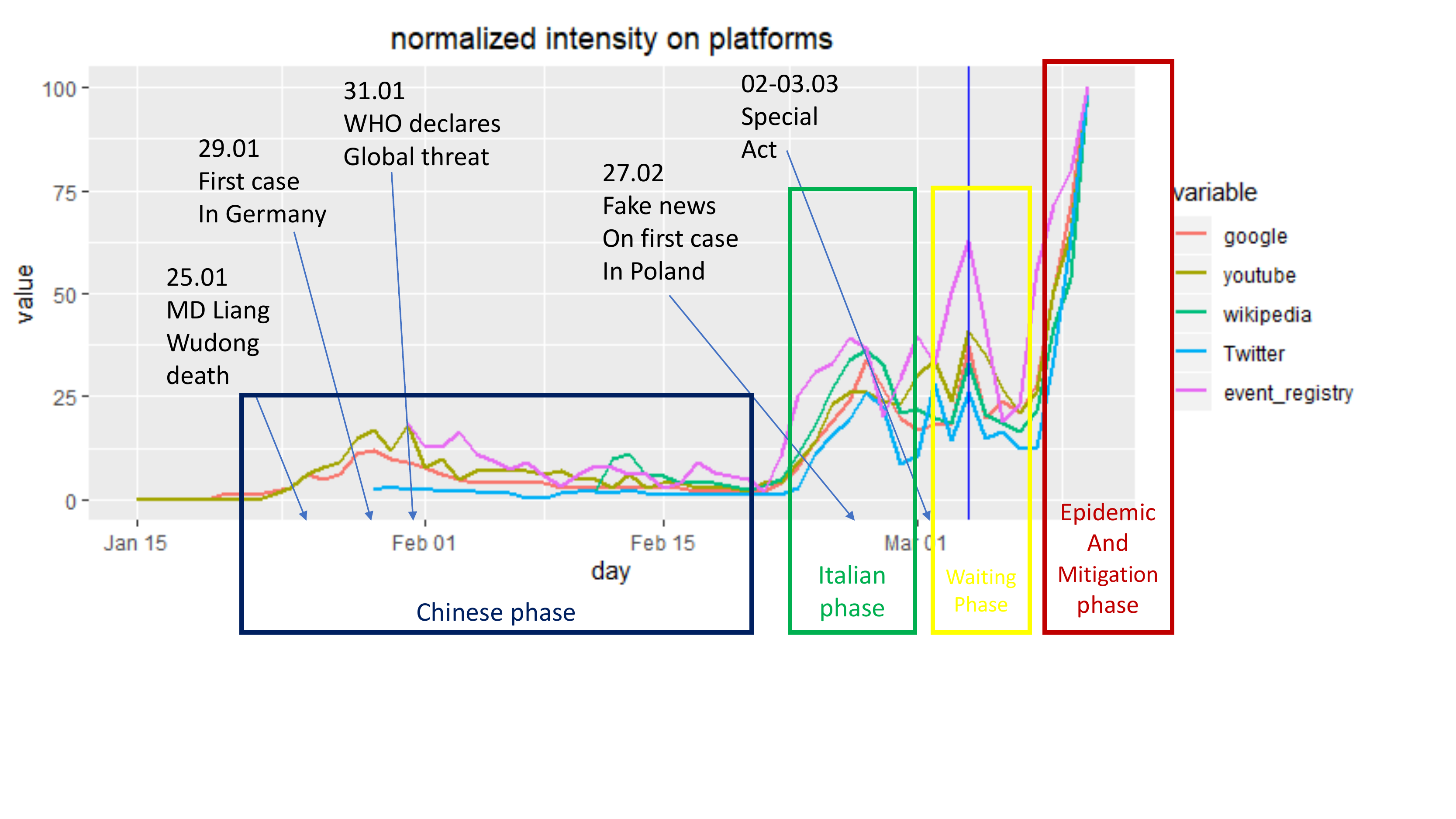}

\caption{\label{time_line} \foreignlanguage{english}{The intensity of topic
Koronawirus on various media platforms during 15.01-11.03.2020. Time
series were normalized to 100 by maximal value for a given serie.
Disease introduction is marked by the blue vertical line}.}
\end{figure}

To compare the interest on COVID-19 on different Internet platforms
we visualized the available queries and interest measurements as time
series together and marked events important to the Polish public~{[}Fig.
\ref{time_line}{]}. From the time series as well as Google queries
semantics~{[}Fig. \ref{ask_public}{]} and tags/topics in the news,
we observe that geographal proximity of the disease drives a lot of
interest (e.g. outbreak in Germany). We see that traditional news
agencies (well represented in EventRegistry) as well as Google search
could anticipate and form more distinct interest peaks than social
media platforms. Moreover discussions on social-contain media (YouTube
and Wikipedia) are smother than information providers (news from EventRegistry,
Google). We detected the lags between different platforms for given
topics (e.g news from event registry are ahead of commentatory media
in fake news on possible introduction). It could mean that the social
media disiminate infomation via speading mechanisms as word-of-mouths.
On the other hand, traditional media have jounlists hired to search
and select most interesting topics quickly (\citep{schultz2011medium}).
It would lead to faster respond to events in news agencies (repersented
by EventRegistry) than social media. A stronger lag (1 day) is observed
between EventRegitry (the most ahead of) and Twitter (the most delayed)
{[}Fig. \ref{lagged}{]}.

We can distinglish phases of interest in Poland (\citep{jarynowski2020percepcja}):

- Chinese phase (COVID-19 emerges in Asia (\citep{strzelecki2020infodemiological})); 

- Itialian phase (second wave of COVID-19 (\citep{strzelecki2020second}));

- Waiting phase (awearness building and waiting for a first case in
Poland);

- Epidemic and Mitigation phase (\citep{inter_ogniska}).

Moreover, Twitter (15\% penetration rate among Polish Internet users
(\citep{pbi_instagram})) has significantly different time pattern
than other analyzed platforms. In particular, it does not have China-related
phase at all and it has at least 3-fold faster growth rate than other
media in epidemic and mitigation phase.

\begin{figure}
\includegraphics[width=0.7\textwidth]{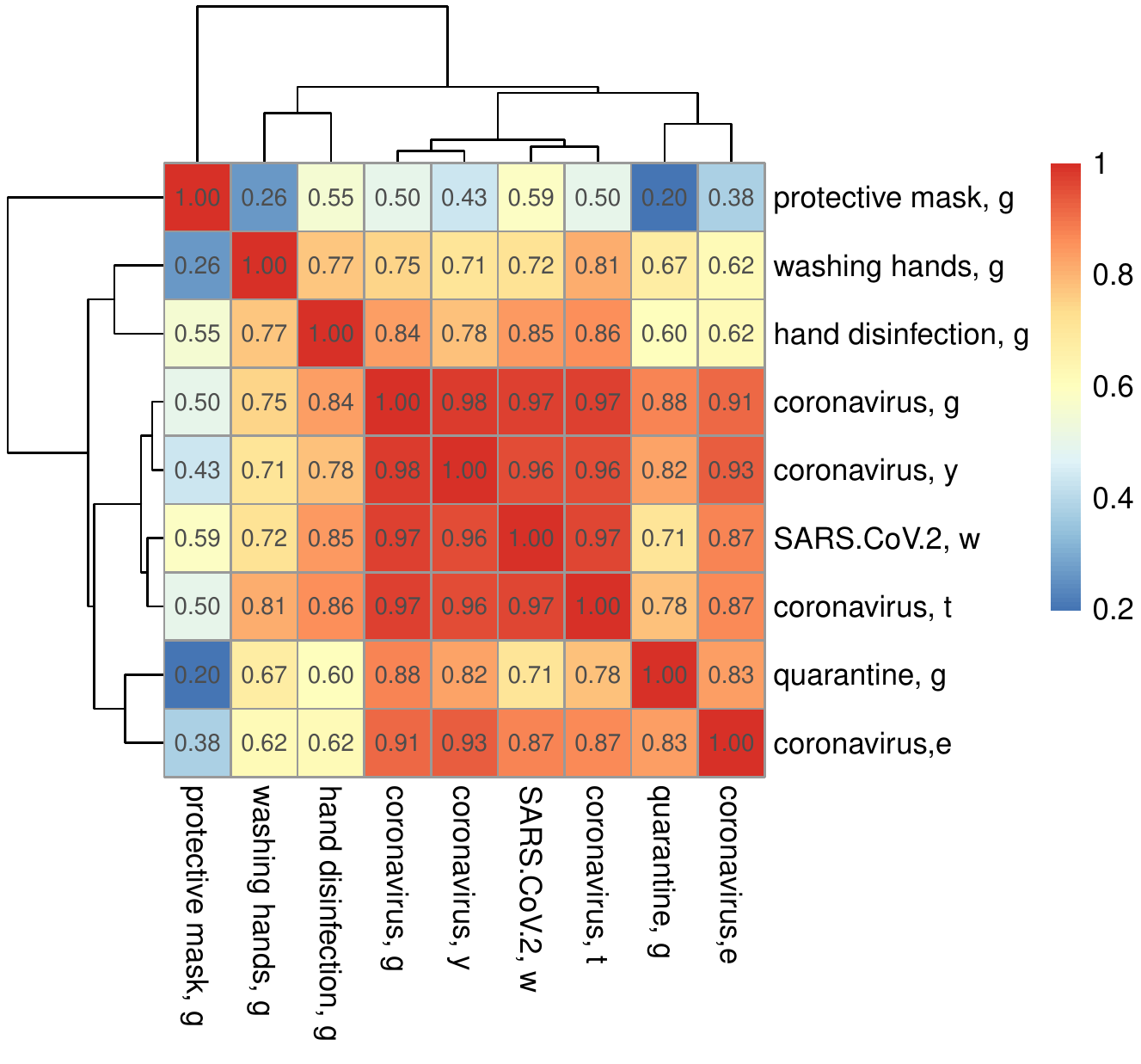}

\caption{\label{corr} \foreignlanguage{english}{The Pearson's correlation
matrix and corresponding hierarhical clustering for the terms ``Koronawirus''/''Covronavius''
and related epidemiological queries on various media platforms} (g~–~Google,
w~–~Wikipedia, y~–~Youtube, t~–~Twitter, e – EventRegistry).
With the significance level of $5\%$ all correlations were significant
except the pair ``antiviral mask, g''/``quarantine, g''. Colorcode
corresponds to correlation strenght.}
\end{figure}

To quantify the correlation in interest on different platforms we
calculated Pearson correlations coefficients between time series~{[}Fig.~\ref{corr}{]}.
All measured intensities are positively correlated. \emph{Protective
mask} as a signal of fear/perception of risk is less correlated with
other variables more related to information needs. \emph{Protective
mask} and \emph{washing hands} is much less correlated than \emph{protective
mask} and \emph{hand disinfection}, which could mean, that people
search for professianal solutions rather thans simple and practical
ones. Moreover, high amounts of searching terms as\emph{ antiviral
mask} (there is no such a medical term) is suggesting that people
are searching colloquial meaning of items (\citep{jarynowski2020percepcja}).

\section*{Conclusions}

\selectlanguage{english}%
In the face of the COVID-19 pandemic, there is an unprecedented flood
of information (information noise) and our task was to extract important
events and features from the most widely range of Internet media in
Poland. The knowledge of quantitative characteristics of the \textquotedbl Coronavirus\textquotedbl{}
perception in Poland is an important prerequisite for a proper crisis
management\foreignlanguage{australian}{~(\citet{trzos2017specyfika})}
such a design of protection policies for risk management and adequate
education of citizens by the stakeholders identified in Poland. For
example, at present, government mitigation programs or hygiene education
principles published on Twitter fall (to the great extend) into the
information bubble or echo chambers (\citep{baumann2020modeling})
of supporters of the ruling party\foreignlanguage{australian}{~{[}Fig.
\ref{net_twi}{]}.} Lack of interest in general public (\citep{marketing,strzelecki2020infodemiological})
could be associated with a low epidemiological awareness of average
Poles, especially when COVID-19 pandemic is massively discussed in
traditional and social media by a relatively small but loud group
of people. 

Before official disease introduction we observed in Poland two information
phases related to outbreaks in China (the end of January) and in Italy
(the second half of February) and one commentary and update phase
in social media in the beginning of March related to, among others,
the special act on COVID-19 mitigation {[}Fig. \ref{time_line}{]}.
Information media (Wikipedia and Google) do not display the third
phase, because probably the awareness about the virus and the knowledge
on the disease has already saturated, and people are interested mainly
in the current update (new infomation on COVID-19) on Twitter, Youtube,
or other electronic media. After the disease introduction to Poland,
there is a peak on the introduction day and fast growth of interest
in epidemic and mitigation phase~{[}Fig.~\ref{time_line}{]}. 
Under conditions of a market-consumer society, private goals of individuals
come into conflict with responsibility of the whole society or community.
For example, economic consequence of supply and demand lead to price
increases of medical devices (\citep{ceneo}).

Due to the data availability, only by analyzing Twitter we have the
full control over the methodology and research techniques. The most
difficult is to access the Facebook data, despite the highest population
penetration and the largest reach in Poland (\citep{media}). Facebook
does not allow automated analysis \foreignlanguage{australian}{(\citet{Face})
}for the public and we can only rely on manual work-intensive research
of commercial companies\foreignlanguage{australian}{ (\citet{sot_T,sot_F}),
}whose research methodology may differ from scientific standards.
However, there are attempts to use available Facebook Ads campaigns
in context to ``Coroanavirus'' \citep{mejova2020advertisers}, which
is available. 

In order to prepare and manage the crisis in an optimal way, a deeper
perception analysis in the form of reliable quantitative and qualitative
analysis is required. Especially, according to the results of empirical
research\foreignlanguage{australian}{~(\citet{taranowicz2010zdrowie}),
}society expects institutional activities and in the event of an epidemic,
it is the \textquotedbl state (...) that is responsible for the poor
health of the population\textquotedbl{} \foreignlanguage{australian}{(\citet{taranowicz2010zdrowie})}.\foreignlanguage{australian}{
Perhaps one of the reasons the Chinese have been so successful in
controlling the spread of the infection is that social media like
WeChat (\citet{wang2020wechat,zhang2020recommended}),} or Internet
forums\foreignlanguage{australian}{ (\citet{liu2018analyzing})} were
analyzed by algorithms\foreignlanguage{australian}{ (\citet{lu2019beyond,paul2011you,salathe2012digital})
with a goal to mitigate the spread of COVID-19. }Combining the behavioral
changes detected via social media analysis with the detailed information
on human mobility via e.g. mobile phone tracking (\citet{schneider2013unravelling,brockmann2006scaling,gonzalez2008understanding}),
sophisticated computational models of infectious disease spread could
be implemented (\citet{belik2011natural,vespignani2012modelling,maier2020effective,hufnagel2004forecast,ferguson2020impact,prasse2020network})
allowing to simulate various scenarios and assess possible human and
economic losses. Furthermore, impact of public information campaigns
could be measured by internal surveys of public opinion. In addition,
Internet media analysis could fill gaps in socio-medical research
on collective actions during an important public health disruption
event such as infectious diseases\foreignlanguage{australian}{ (\citet{jarynowski2019kosztaASf}).}

\selectlanguage{australian}%
Acknowledgments: we thanks PNFN (2019-21), NCN (2016/22/E/HS2/00034),
and FU Berlin (FU AvH: 08166500) for partial finanacial support and
Łukasz Krzowski, Daniel Płatek, Ireneusz Skawina, Andrzej Buda, and
Marcus Doherr for fruitful discussions.

\section*{References}

\bibliographystyle{unsrtnat}
\addcontentsline{toc}{section}{\refname}\bibliography{bibl_new2}

\section*{}
\end{document}

%% file: img_dir/table211.tex
\begin{adjustbox}{width=\columnwidth,center}
\begin{tabular}{ cc }
  \centering
\begin{tabular}{clll}
\toprule
Rank & item & count & item English\\
\midrule
1&polska&32266&Poland\\
2&rzad	&	22500&	Government\\
3&przypadek	&	21868	&	case	\\
4&osoba	&	18723&	person\\
5&polski	&	18482&	Polish\\
6&wszystek	&	17924&	all\\
7&móc	&	17908&	can\\
8&zdrowie	&	17476&	health\\
9&wlochy	&	14113&	Italy	\\
10&czlowiek	&	13681&	men	\\
11&wszyscy	&	13586&	all\\
12&minister	&	13079&	Minister\\
13&test	&	13079	&test\\
14&pis	&	12810	&Law and Justice\\
15&ludzie	&	12602	&people\\
\bottomrule
\end{tabular} &
\begin{tabular}{clll}
\toprule
Rank & item & count & item English\\
\midrule
16&	informacja	&	12248	&	information\\
17&	szpital	&	12217	&	hospital\\
18&	szkola	&	12168	&	school\\
19&	chiny	&	12153	&	China\\
20&	zarazic	&	12127	&	to infect\\
21&	epidemia	&	12117	&	epidemic	\\
22&	wirus	&	11349	&	virus\\
23&	walka	&	10904	&	fight\\
24&	zakazic	&	10888	&	to infect\\
25&	kwarantanna	&	10552	&	quarantaine\\
26&	potwierdzic	&	10423	&	to confirm\\
27&	dzien	&	10393	&	day\\
28&	polak	&	10341	&	Pole	\\
29&	panstwo	&	9769	&	country\\
30&	zamknac	&	9713	&	to close\\
\bottomrule
\end{tabular}\\
\end{tabular}
\end{adjustbox}

%% file: med_arch2_wit.bbl
\begin{thebibliography}{69}
\providecommand{\natexlab}[1]{#1}
\providecommand{\url}[1]{\texttt{#1}}
\expandafter\ifx\csname urlstyle\endcsname\relax
  \providecommand{\doi}[1]{doi: #1}\else
  \providecommand{\doi}{doi: \begingroup \urlstyle{rm}\Url}\fi

\bibitem[IBRIS(2020)]{ibris}
IBRIS.
\newblock Koronawirus ponad polityka.
\newblock
  \url{https://wiadomosci.onet.pl/tylko-w-onecie/wybory-prezydenckie-2020-sondaz-andrzej-duda-prowadzi-potrzebna-ii-tura/cmk8ssw},
  2020.
\newblock Accessed: 2020-03-04.

\bibitem[OKO(2020)]{oko_2020}
OKO.
\newblock Sondaż ipsos o koronawirusie. dwie trzecie widzi duże zagrożenie,
  ale nie panikujemy.
\newblock \url{https://oko.press/sondaz-ipsos-o-koronawirusie/}, 2020.
\newblock Accessed: 2020-03-12.

\bibitem[Ginsberg et~al.(2009)Ginsberg, Mohebbi, Patel, Brammer, Smolinski, and
  Brilliant]{ginsberg2009detecting}
Jeremy Ginsberg, Matthew~H Mohebbi, Rajan~S Patel, Lynnette Brammer, Mark~S
  Smolinski, and Larry Brilliant.
\newblock Detecting influenza epidemics using search engine query data.
\newblock \emph{Nature}, 457\penalty0 (7232):\penalty0 1012--1014, 2009.

\bibitem[Lu et~al.(2018)Lu, Hou, Baltrusaitis, Shah, Leskovec, Hawkins,
  Brownstein, Conidi, Gunn, Gray, et~al.]{lu2018accurate}
Fred~Sun Lu, Suqin Hou, Kristin Baltrusaitis, Manan Shah, Jure Leskovec, Jared
  Hawkins, John Brownstein, Giuseppe Conidi, Julia Gunn, Josh Gray, et~al.
\newblock Accurate influenza monitoring and forecasting using novel internet
  data streams: a case study in the boston metropolis.
\newblock \emph{JMIR public health and surveillance}, 4\penalty0 (1):\penalty0
  e4, 2018.

\bibitem[Joshi et~al.(2020)Joshi, Sparks, McHugh, Karimi, Paris, and
  MacIntyre]{joshi2020harnessing}
Aditya Joshi, Ross Sparks, James McHugh, Sarvnaz Karimi, Cecile Paris, and
  C~Raina MacIntyre.
\newblock Harnessing tweets for early detection of an acute disease event.
\newblock \emph{Epidemiology}, 31\penalty0 (1):\penalty0 90--97, 2020.

\bibitem[Kata(2012)]{kata2012anti}
Anna Kata.
\newblock Anti-vaccine activists, web 2.0, and the postmodern paradigm--an
  overview of tactics and tropes used online by the anti-vaccination movement.
\newblock \emph{Vaccine}, 30\penalty0 (25):\penalty0 3778--3789, 2012.

\bibitem[Kassan(2020)]{natasza}
Natasha Kassan.
\newblock Disinformation and coronavirus.
\newblock
  \url{https://www.lowyinstitute.org/the-interpreter/disinformation-and-coronavirus},
  2020.

\bibitem[Guardian(2020)]{guardian}
Guardian.
\newblock Italian minister tries to calm coronavirus panic and attacks
  profiteers.
\newblock
  \url{https://www.theguardian.com/world/2020/feb/27/italian-minister-tries-to-calm-coronavirus-panic-and-attacks-profiteers},
  2020.
\newblock Accessed: 2020-03-04.

\bibitem[Jacobs(2017)]{lightfoot2017political}
Sean Jacobs.
\newblock Political propaganda spread through social bots.
\newblock \emph{Media, Culture, \& Global Politics}, pages 1--22, 2017.

\bibitem[Gonsalves and Staley(2014)]{gonsalves2014panic}
Gregg Gonsalves and Peter Staley.
\newblock Panic, paranoia, and public health—the aids epidemic's lessons for
  ebola.
\newblock \emph{New England Journal of Medicine}, 371\penalty0 (25):\penalty0
  2348--2349, 2014.

\bibitem[Perra et~al.(2011)Perra, Balcan, Gon{\c{c}}alves, and
  Vespignani]{perra2011towards}
Nicola Perra, Duygu Balcan, Bruno Gon{\c{c}}alves, and Alessandro Vespignani.
\newblock Towards a characterization of behavior-disease models.
\newblock \emph{PloS one}, 6\penalty0 (8), 2011.

\bibitem[Fenichel et~al.(2011)Fenichel, Castillo-Chavez, Ceddia, Chowell,
  Parra, Hickling, Holloway, Horan, Morin, Perrings,
  et~al.]{fenichel2011adaptive}
Eli~P Fenichel, Carlos Castillo-Chavez, M~Graziano Ceddia, Gerardo Chowell,
  Paula A~Gonzalez Parra, Graham~J Hickling, Garth Holloway, Richard Horan,
  Benjamin Morin, Charles Perrings, et~al.
\newblock Adaptive human behavior in epidemiological models.
\newblock \emph{Proceedings of the National Academy of Sciences}, 108\penalty0
  (15):\penalty0 6306--6311, 2011.

\bibitem[Wang et~al.(2015)Wang, Andrews, Wu, Wang, and Bauch]{wang2015coupled}
Zhen Wang, Michael~A Andrews, Zhi-Xi Wu, Lin Wang, and Chris~T Bauch.
\newblock Coupled disease--behavior dynamics on complex networks: A review.
\newblock \emph{Physics of life reviews}, 15:\penalty0 1--29, 2015.

\bibitem[Samaras et~al.(2020)Samaras, Garc{\'\i}a-Barriocanal, and
  Sicilia]{samaras2020syndromic}
Loukas Samaras, Elena Garc{\'\i}a-Barriocanal, and Miguel-Angel Sicilia.
\newblock Syndromic surveillance using web data: a systematic review.
\newblock In \emph{Innovation in Health Informatics}, pages 39--77. Elsevier,
  2020.

\bibitem[Nuti et~al.(2014)Nuti, Wayda, Ranasinghe, Wang, Dreyer, Chen, and
  Murugiah]{nuti2014use}
Sudhakar~V Nuti, Brian Wayda, Isuru Ranasinghe, Sisi Wang, Rachel~P Dreyer,
  Serene~I Chen, and Karthik Murugiah.
\newblock The use of google trends in health care research: a systematic
  review.
\newblock \emph{PloS one}, 9\penalty0 (10), 2014.

\bibitem[Jarynowski et~al.(2020)Jarynowski, Wojta-Kempa, and
  Belik]{jarynowski2020percepcja}
Andrzej Jarynowski, Monika Wojta-Kempa, and Vitaly Belik.
\newblock Percepcja „koronawirusa” w polskim internecie do czasu
  potwierdzenia pierwszego przypadku zakażenia sars-cov-2 w polsce.
\newblock \emph{accpeted: Pielegniarstwo i Zdrowie Publiczne}, 2020.

\bibitem[PBI(2020)]{pbi_20}
PBI.
\newblock Polski internet w styczniu 2020.
\newblock
  \url{http://pbi.org.pl/badanie-gemius-pbi/polski-internet-w-styczniu-2020/},
  2020.
\newblock Accessed: 2020-03-04.

\bibitem[PBI(2019)]{pbi_19}
PBI.
\newblock Polski internet w styczniu 2019.
\newblock
  \url{http://pbi.org.pl/badanie-gemius-pbi/polski-internet-w-styczniu-2019/},
  2019.
\newblock Accessed: 2020-03-04.

\bibitem[Tana{\'s} et~al.(2017)Tana{\'s}, Kamieniecki, Bochenek, and
  Lange]{tanas2017raport}
M~Tana{\'s}, W~Kamieniecki, M~Bochenek, and R~Lange.
\newblock Raport z badania nastolatki 3.0.
\newblock \emph{Warszawa, NASK}, 2017.

\bibitem[Jarynowski and Belik(2018)]{jarynowski2018choroby}
Andrzej Jarynowski and Vitaly Belik.
\newblock Choroby przenoszone drog{\k{a}} p{\l}ciow{\k{a}} w dobie internetu i
  e-zdrowia: kalkulatory ryzyka.
\newblock \emph{Krak{\'o}w: Biblioteka Jagiello{\'n}ska}, 2018.

\bibitem[Jarynowski et~al.(2014)Jarynowski, Buda, and Nyczka]{Obliczeniowe}
Andrzej Jarynowski, Andrzej Buda, and Piotr Nyczka.
\newblock \emph{Obliczeniowe nauki spoleczne w Praktycze}.
\newblock WN:Glogow, 2014.

\bibitem[Wyborcza(2020)]{wyborcza}
Gazeta Wyborcza.
\newblock Chybicka: "w Łodzi potwierdzono koronawirusa".
\newblock
  \url{https://wiadomosci.gazeta.pl/wiadomosci/7,114884,25741826,chybicka-w-lodzi-potwierdzono-koronawirusa-minister-zdrowia.html},
  2020.
\newblock Accessed: 2020-03-12.

\bibitem[Lai et~al.(2020)Lai, Bogoch, Ruktanonchai, Watts, Li, Yu, Lv, Yang,
  Yu, Khan, et~al.]{lai2020assessing}
Shengjie Lai, Isaac Bogoch, Nick Ruktanonchai, Alexander Watts, Yu~Li, Jianzing
  Yu, Xin Lv, Weizhong Yang, Hongjie Yu, Kamran Khan, et~al.
\newblock Assessing spread risk of wuhan novel coronavirus within and beyond
  china, january-april 2020: a travel network-based modelling study.
\newblock \emph{medRxiv}, 2020.

\bibitem[Interdisciplinary(2020{\natexlab{a}})]{inter}
Interdisciplinary.
\newblock Kiedy 2019-ncov trafi do polski?
\newblock
  \url{http://interdisciplinary-research.eu/kiedy-2019n-cov-trafi-do-polski},
  2020{\natexlab{a}}.
\newblock Accessed: 2020-03-04.

\bibitem[Maziarz et~al.(2016)Maziarz, Piasecki, Rudnicka, Szpakowicz, and
  K{\k{e}}dzia]{maziarz2016plwordnet}
Marek Maziarz, Maciej Piasecki, Ewa Rudnicka, Stan Szpakowicz, and Pawe{\l}
  K{\k{e}}dzia.
\newblock plwordnet 3.0--a comprehensive lexical-semantic resource.
\newblock In \emph{Proceedings of COLING 2016, the 26th International
  Conference on Computational Linguistics: Technical Papers}, pages 2259--2268,
  2016.

\bibitem[AnswearPublic(2020)]{ask_public}
AnswearPublic.
\newblock Answear the public.
\newblock \url{https://answerthepublic.com}, 2020.
\newblock Accessed: 2020-03-04.

\bibitem[PBI(2018{\natexlab{a}})]{pbi_wiki}
PBI.
\newblock Wikipedia i jej użytkownicy.
\newblock
  \url{http://pbi.org.pl/wp-content/uploads/2017/09/2017-09-26-Wikipedia_analiza.pdf},
  2018{\natexlab{a}}.
\newblock Accessed: 2020-03-12.

\bibitem[Wikipedia(2020{\natexlab{a}})]{wiki_SARS-CoV-2}
Wikipedia.
\newblock Sars-cov-2.
\newblock \url{https://pl.wikipedia.org/wiki/SARS-CoV-2}, 2020{\natexlab{a}}.
\newblock Accessed: 2020-03-04.

\bibitem[Wikipedia(2020{\natexlab{b}})]{wiki_szerzenie}
Wikipedia.
\newblock Szerzenie sie zakazen wirusem sars-cov-2.
\newblock
  \url{https://pl.wikipedia.org/wiki/Szerzenie_się_zakazen_wirusem_SARS-CoV-2},
  2020{\natexlab{b}}.
\newblock Accessed: 2020-03-04.

\bibitem[EventRegistry(2020)]{Event}
EventRegistry.
\newblock Event registry.
\newblock \url{https://www.eventregistry.org}, 2020.
\newblock Accessed: 2020-03-04.

\bibitem[Sotrender(2019)]{sot_T}
Sotrender.
\newblock Twitter w polsce – podsumowanie 2018 roku.
\newblock
  \url{https://www.sotrender.com/blog/pl/2019/02/twitter-w-polsce-podsumowanie-2018-roku-infografika/},
  2019.
\newblock Accessed: 2020-03-04.

\bibitem[Sejm(2020)]{sejm}
Sejm.
\newblock Rzadowy projekt ustawy o szczegolnych rozwiazaniach zwiazanych z
  zapobieganiem, przeciwdzialaniem i zwalczaniem covid-19.
\newblock
  \url{https://www.sejm.gov.pl/Sejm9.nsf/PrzebiegProc.xsp?id=016EAA75EDD551EBC125851E0077C1C2},
  2020.
\newblock Accessed: 2020-03-04.

\bibitem[Wasserman et~al.(1994)Wasserman, Faust, et~al.]{wasserman1994social}
Stanley Wasserman, Katherine Faust, et~al.
\newblock \emph{Social network analysis: Methods and applications}, volume~8.
\newblock Cambridge university press, 1994.

\bibitem[Jarynowski et~al.(2019{\natexlab{a}})Jarynowski, Paradowski, and
  Buda]{modelowanie}
Andrzej Jarynowski, Michal~B Paradowski, and Andrzej Buda.
\newblock Modelling communities and populations: an introduction to
  computational social science.
\newblock \emph{Studia metodologiczne}, 39:\penalty0 117--139,
  2019{\natexlab{a}}.

\bibitem[Blondel et~al.(2008)Blondel, Guillaume, Lambiotte, and
  Lefebvre]{blondel2008fast}
Vincent~D Blondel, Jean-Loup Guillaume, Renaud Lambiotte, and Etienne Lefebvre.
\newblock Fast unfolding of communities in large networks.
\newblock \emph{Journal of statistical mechanics: theory and experiment},
  2008\penalty0 (10):\penalty0 P10008, 2008.

\bibitem[OKO(2019)]{oko}
OKO.
\newblock Proba wplyniecia na wyniki wyborow? dwie siatki patriotycznych trolli
  wspieraly konfederacje.
\newblock
  \url{https://oko.press/proba-wplyniecia-na-wyniki-wyborow-dwie-siatki-patriotycznych-trolli-wspieraly-konfederacje/},
  2019.
\newblock Accessed: 2020-03-04.

\bibitem[Jarynowski et~al.(2019{\natexlab{b}})Jarynowski, Platek, Krzowski,
  Gerylovich, and Belik]{jarynowski2019african}
Andrzej Jarynowski, Daniel Platek, {\L}ukasz Krzowski, Anton Gerylovich, and
  Vitaly Belik.
\newblock African swine fever-potential biological warfare threat.
\newblock Technical report, EasyChair, 2019{\natexlab{b}}.

\bibitem[Youtube(2020)]{Youtube}
Youtube.
\newblock Koronawirus.
\newblock \url{https://www.youtube.com/}, 2020.
\newblock Accessed: 2020-03-04.

\bibitem[Sotrender(2020)]{sot_F}
Sotrender.
\newblock Facebook trends styczen 2020 – burzliwych 31 dni nowego roku.
\newblock
  \url{https://www.sotrender.com/blog/pl/2020/02/facebook-trends-styczen-2020-burzliwych-31-dni-nowego-roku/},
  2020.
\newblock Accessed: 2020-03-04.

\bibitem[PBI(2018{\natexlab{b}})]{pbi_instagram}
PBI.
\newblock Instagram i jego polscy użytkownicy.
\newblock
  \url{http://pbi.org.pl/wp-content/uploads/2018/07/2018-07-Instagram.pdf},
  2018{\natexlab{b}}.
\newblock Accessed: 2020-03-12.

\bibitem[Media@jesion(2020)]{blog}
Media@jesion.
\newblock Pytania i odpowiedzi dotyczące covid-19.
\newblock
  \url{https://medium.com/@jesion/dlaczego-powinniśmy-się-przejmować-często-zadawane-pytania-i-odpowiedzi-dotyczące-covid-19},
  2020.
\newblock Accessed: 2020-03-12.

\bibitem[Schultz et~al.(2011)Schultz, Utz, and G{\"o}ritz]{schultz2011medium}
Friederike Schultz, Sonja Utz, and Anja G{\"o}ritz.
\newblock Is the medium the message? perceptions of and reactions to crisis
  communication via twitter, blogs and traditional media.
\newblock \emph{Public relations review}, 37\penalty0 (1):\penalty0 20--27,
  2011.

\bibitem[Strzelecki(2020{\natexlab{a}})]{strzelecki2020infodemiological}
Artur Strzelecki.
\newblock Infodemiological study using google trends on coronavirus epidemic in
  wuhan, china.
\newblock \emph{arXiv preprint arXiv:2001.11021}, 2020{\natexlab{a}}.

\bibitem[Strzelecki(2020{\natexlab{b}})]{strzelecki2020second}
Artur Strzelecki.
\newblock The second worldwide wave of interest in coronavirus since the
  covid-19 outbreaks in south korea, italy and iran: A google trends study.
\newblock \emph{arXiv preprint arXiv:2003.10998}, 2020{\natexlab{b}}.

\bibitem[Interdisciplinary(2020{\natexlab{b}})]{inter_ogniska}
Interdisciplinary.
\newblock Sars-cov-2 opracowanie pierwszych ognisk.
\newblock
  \url{http://interdisciplinary-research.eu/sars-cov-2-opracowanie-pierwszych-ognisk},
  2020{\natexlab{b}}.
\newblock Accessed: 2020-03-16.

\bibitem[Trzos et~al.(2017)Trzos, Krzowski, and D{\l}ugosz]{trzos2017specyfika}
Arkadiusz Trzos, {\L}ukasz Krzowski, and Katarzyna D{\l}ugosz.
\newblock Specyfika dzia{\l}a{\'n} ratownictwa medycznego.
\newblock \emph{Na Ratunek}, 4:\penalty0 17, 2017.

\bibitem[Baumann et~al.(2020)Baumann, Lorenz-Spreen, Sokolov, and
  Starnini]{baumann2020modeling}
Fabian Baumann, Philipp Lorenz-Spreen, Igor~M Sokolov, and Michele Starnini.
\newblock Modeling echo chambers and polarization dynamics in social networks.
\newblock \emph{Physical Review Letters}, 124\penalty0 (4):\penalty0 048301,
  2020.

\bibitem[Widzialni(2020)]{marketing}
Widzialni.
\newblock Marketing wirusowy koronawirusowa.
\newblock \url{https://www.widzialni.pl/blog/marketing-wirusowy-koronawirusa/},
  2020.
\newblock Accessed: 2020-03-04.

\bibitem[Ceneo(2020)]{ceneo}
Ceneo.
\newblock Koronawirus atakuje sklepy. boom na maski i… nie tylko.
\newblock
  \url{https://subiektywnieofinansach.pl/koronawirus-atakuje-a-w-sklepach-boom-na-maski-i-nie-tylko-w-trakcie-kompletowania-zamowienia-cena-skoczyla-czterokrotnie/},
  2020.
\newblock Accessed: 2020-03-04.

\bibitem[Mobirank(2018)]{media}
Mobirank.
\newblock Social media users in poland.
\newblock
  \url{https://mobirank.pl/2018/12/04/liczba-uzytkownikow-facebooka-instagrama-i-messengera-w-polsce-11-2018/},
  2018.
\newblock Accessed: 2020-03-04.

\bibitem[Facebook(2020)]{Face}
Facebook.
\newblock Facebook for developers. documentation.
\newblock \url{https://developers.facebook.com/docs}, 2020.
\newblock Accessed: 2020-03-04.

\bibitem[Mejova and Kalimeri(2020)]{mejova2020advertisers}
Yelena Mejova and Kyriaki Kalimeri.
\newblock Advertisers jump on coronavirus bandwagon: Politics, news, and
  business.
\newblock \emph{arXiv preprint arXiv:2003.00923}, 2020.

\bibitem[Taranowicz(2010)]{taranowicz2010zdrowie}
Iwona Taranowicz.
\newblock Zdrowie i sposoby radzenia sobie z jego zagro{\.z}eniami.
\newblock \emph{Analiza socjologiczna. Wroc{\l}aw: Oficyna Wydawnicza
  Arboretum}, 2010.

\bibitem[Wang et~al.(2020)Wang, Wang, Zhang, Li, Jia, and Dang]{wang2020wechat}
Wenjun Wang, Yikai Wang, Xin Zhang, Yaping Li, Xiaoli Jia, and Shuangsuo Dang.
\newblock Wechat, a chinese social media, may early detect the sars-cov-2
  outbreak in 2019.
\newblock \emph{medRxiv}, 2020.

\bibitem[Zhang et~al.(2020)Zhang, Wu, Zhao, and Zhang]{zhang2020recommended}
Jun Zhang, Weili Wu, Xin Zhao, and Wei Zhang.
\newblock Recommended psychological crisis intervention response to the 2019
  novel coronavirus pneumonia outbreak in china: a model of west china
  hospital.
\newblock \emph{Precision Clinical Medicine}, 2020.

\bibitem[Liu and Lu(2018)]{liu2018analyzing}
Chuchu Liu and Xin Lu.
\newblock Analyzing hidden populations online: topic, emotion, and social
  network of hiv-related users in the largest chinese online community.
\newblock \emph{BMC medical informatics and decision making}, 18\penalty0
  (1):\penalty0 2, 2018.

\bibitem[Lu et~al.(2019)Lu, Qin, Holme, Meng, Hu, Liljeros, and
  Allon]{lu2019beyond}
Xin Lu, Shuo Qin, Petter Holme, Fanhui Meng, Yanqing Hu, Fredrik Liljeros, and
  Gad Allon.
\newblock Beyond the coverage of information spreading: Analytical and
  empirical evidence of re-exposure in large-scale online social networks.
\newblock \emph{arXiv preprint arXiv:1907.12389}, 2019.

\bibitem[Paul and Dredze(2011)]{paul2011you}
Michael~J Paul and Mark Dredze.
\newblock You are what you tweet: Analyzing twitter for public health.
\newblock In \emph{Fifth International AAAI Conference on Weblogs and Social
  Media}, 2011.

\bibitem[Salathe et~al.(2012)Salathe, Bengtsson, Bodnar, Brewer, Brownstein,
  Buckee, Campbell, Cattuto, Khandelwal, Mabry, et~al.]{salathe2012digital}
Marcel Salathe, Linus Bengtsson, Todd~J Bodnar, Devon~D Brewer, John~S
  Brownstein, Caroline Buckee, Ellsworth~M Campbell, Ciro Cattuto, Shashank
  Khandelwal, Patricia~L Mabry, et~al.
\newblock Digital epidemiology.
\newblock \emph{PLoS computational biology}, 8\penalty0 (7), 2012.

\bibitem[Schneider et~al.(2013)Schneider, Belik, Couronn{\'e}, Smoreda, and
  Gonz{\'a}lez]{schneider2013unravelling}
Christian~M Schneider, Vitaly Belik, Thomas Couronn{\'e}, Zbigniew Smoreda, and
  Marta~C Gonz{\'a}lez.
\newblock Unravelling daily human mobility motifs.
\newblock \emph{Journal of The Royal Society Interface}, 10\penalty0
  (84):\penalty0 20130246, 2013.

\bibitem[Brockmann et~al.(2006)Brockmann, Hufnagel, and
  Geisel]{brockmann2006scaling}
Dirk Brockmann, Lars Hufnagel, and Theo Geisel.
\newblock The scaling laws of human travel.
\newblock \emph{Nature}, 439\penalty0 (7075):\penalty0 462--465, 2006.

\bibitem[Gonzalez et~al.(2008)Gonzalez, Hidalgo, and
  Barabasi]{gonzalez2008understanding}
Marta~C Gonzalez, Cesar~A Hidalgo, and Albert-Laszlo Barabasi.
\newblock Understanding individual human mobility patterns.
\newblock \emph{nature}, 453\penalty0 (7196):\penalty0 779--782, 2008.

\bibitem[Belik et~al.(2011)Belik, Geisel, and Brockmann]{belik2011natural}
Vitaly Belik, Theo Geisel, and Dirk Brockmann.
\newblock Natural human mobility patterns and spatial spread of infectious
  diseases.
\newblock \emph{Physical Review X}, 1\penalty0 (1):\penalty0 011001, 2011.

\bibitem[Vespignani(2012)]{vespignani2012modelling}
Alessandro Vespignani.
\newblock Modelling dynamical processes in complex socio-technical systems.
\newblock \emph{Nature physics}, 8\penalty0 (1):\penalty0 32--39, 2012.

\bibitem[Maier and Brockmann(2020)]{maier2020effective}
Benjamin~F Maier and Dirk Brockmann.
\newblock Effective containment explains sub-exponential growth in confirmed
  cases of recent covid-19 outbreak in mainland china.
\newblock \emph{arXiv preprint arXiv:2002.07572}, 2020.

\bibitem[Hufnagel et~al.(2004)Hufnagel, Brockmann, and
  Geisel]{hufnagel2004forecast}
Lars Hufnagel, Dirk Brockmann, and Theo Geisel.
\newblock Forecast and control of epidemics in a globalized world.
\newblock \emph{Proceedings of the National Academy of Sciences}, 101\penalty0
  (42):\penalty0 15124--15129, 2004.

\bibitem[Ferguson et~al.(2020)Ferguson, Laydon, Nedjati-Gilani, Imai, Ainslie,
  Baguelin, Bhatia, Boonyasiri, Cucunub{\'a}, Cuomo-Dannenburg,
  et~al.]{ferguson2020impact}
Neil~M Ferguson, Daniel Laydon, Gemma Nedjati-Gilani, Natsuko Imai, Kylie
  Ainslie, Marc Baguelin, Sangeeta Bhatia, Adhiratha Boonyasiri, Zulma
  Cucunub{\'a}, Gina Cuomo-Dannenburg, et~al.
\newblock Impact of non-pharmaceutical interventions (npis) to reduce covid-19
  mortality and healthcare demand.
\newblock \emph{Imperial College, London doi.org/10.25561/77482}, 2020.

\bibitem[Prasse et~al.(2020)Prasse, Achterberg, Ma, and
  Van~Mieghem]{prasse2020network}
Bastian Prasse, Massimo~A Achterberg, Long Ma, and Piet Van~Mieghem.
\newblock Network-based prediction of the 2019-ncov epidemic outbreak in the
  chinese province hubei.
\newblock \emph{arXiv preprint arXiv:2002.04482}, 2020.

\bibitem[Jarynowski and Belik(2019)]{jarynowski2019kosztaASf}
Andrzej Jarynowski and Vitaly Belik.
\newblock Analiza kosztow rozprzestrzeniania sie afrykanskiego pomoru swin w
  polsce.
\newblock \emph{Public Health Forum}, V(XIII):\penalty0 72, 2019.

\end{thebibliography}
